\documentclass[sigconf]{acmart}
\usepackage{xcolor}  % Load without options

\definecolor{myred}{RGB}{200,0,0}
\definecolor{mygreen}{RGB}{0,150,0}

\usepackage{multirow}
\usepackage{longtable}
\usepackage{rotating}
\usepackage{setspace}
\usepackage{multirow}
\usepackage{graphicx}

\usepackage{enumitem}

\usepackage{amsmath}
\usepackage{pifont}

\newcommand{\xmark}{\ding{55}}%

\AtBeginDocument{%
  \providecommand\BibTeX{{%
    \normalfont B\kern-0.5em{\scshape i\kern-0.25em b}\kern-0.8em\TeX}}}

\copyrightyear{2025}
\acmYear{2025}
\setcopyright{acmcopyright}
\acmConference[]{}{}{}
\acmConference[ICAIL '25]{International Conference on Artificial Intelligence and Law}{June 16-20}{Chicago, IL}
\title{From Legal Text to Tech Specs: Generative AI’s Interpretation of Consent in Privacy Law}

\author{Aniket Kesari}
\affiliation{%
  \institution{Fordham University}
  \streetaddress{150 W 62nd St, New York, NY 10023}
 \city{New York}
  \country{United States}}
 \email{akesari@fordham.edu}

\author{Travis Breaux}
\affiliation{%
  \institution{Carnegie Mellon University}
  \streetaddress{}
 \city{Pittsburgh}
  \country{USA}}
  \email{tdbreaux@andrew.cmu.edu}

 \author{Tom Norton}
\affiliation{%
  \institution{Fordham University}
  \streetaddress{150 W 62nd St, New York, NY 10023}
 \city{New York}
  \country{United States}}
 \email{tnorton1@law.fordham.edu}

\author{Sarah Santos}
\affiliation{%
  \institution{Carnegie Mellon University}
  \streetaddress{}
 \city{Pittsburgh}
  \country{USA}}
  \email{ssantos@andrew.cmu.edu }

\author{Anmol Singhal}
\affiliation{%
  \institution{Carnegie Mellon University}
  \streetaddress{}
 \city{Pittsburgh}
  \country{USA}}
  \email{anmolsinghal@cmu.edu }
 
\date{September 2024}

\begin{abstract}
    Privacy law and regulation have turned to ``consent'' as the legitimate basis for collecting and processing individuals' data. As governments have rushed to enshrine consent requirements in their privacy laws, such as the California Consumer Privacy Act (CCPA), significant challenges remain in understanding how these legal mandates are operationalized in software. The opaque nature of software development processes further complicates this translation. To address this, we explore the use of Large Language Models (LLMs) in requirements engineering to bridge the gap between legal requirements and technical implementation. This study employs a three-step pipeline that involves using an LLM to classify software use cases for compliance, generating LLM modifications for non-compliant cases, and manually validating these changes against legal standards. Our preliminary findings highlight the potential of LLMs in automating compliance tasks, while also revealing limitations in their reasoning capabilities. By benchmarking LLMs against real-world use cases, this research provides insights into leveraging AI-driven solutions to enhance legal compliance of software.

\end{abstract}



\begin{document}

\maketitle
\pagestyle{empty}

\section{Introduction}
\label{section:intro}
Consent has long been the foundation of privacy rights in both the United States and the European Union. The traditional justification for consent-based frameworks is that consent preserves market mechanisms, consumer choice, and personal control over data. However, privacy scholars have increasingly critiqued these frameworks, arguing that their practical implementation often falls short. Behavioral economics research suggests that consumers, overwhelmed by excessive information, make irrational decisions about their privacy. Others contend that consent mechanisms fail to offer meaningful choice or safeguard against broader societal harms, such as risks to individual data subject's social networks and society at large. These debates are important because the right answer will dictate the eventual interventions – do we just need better notices or more direct regulation of information flows?
    
Missing from this debate is an empirical understanding of how consent is operationalized in real-world applications. The California Consumer Privacy Act (CCPA) requires that privacy notices be ``clear and conspicuous,'' but how do businesses actually interpret and implement this requirement? Legal scholars know that seemingly simple terms such as “freely given” and “informed” can actually be laden with thick substantive nuance. However, it is unclear whether these complexities are reflected in software. This lack of clarity is exacerbated by the lack of transparency in software development processes, which often occur “behind the scenes” and are inaccessible to regulators and the public. Conversations between engineers and legal teams are  proprietary, making it difficult to discern how legal principles translate into software requirements.

These challenges highlight the need for innovative tools to bridge the gap between legal requirements and their technical implementation. In this paper, we explore a novel application of generative Artificial Intelligence (AI), specifically Large Language Models (LLMs), to address this issue. LLMs have shown promise in automating complex tasks across various domains, including requirements engineering—a sub-discipline of computer science focused on defining the expectations and objectives of software systems. Our research investigates whether LLMs can assist in ensuring software compliance with legal frameworks for informed consent. More specifically, we analyze if LLMs can identify non-compliant use cases and suggest actionable modifications.

Our method relies on an assumption about the future of software development: that it will increasingly rely on \textit{generative} approaches. Developers are beginning to use use LLMs to generate and critique requirements~\cite{arora2024advancing}, create user personas~\cite{zhang2023personas, hamalainen2023evaluating}, design artifacts~\cite{eisenreich2024requirements}, and generate code and test cases~\cite{schafer2023empirical}. However, the generative nature of these tools introduces challenges such as hallucinations, sycophancy, and inconsistencies, which pose challenges for conforming generated products to legal and societal expectations. Moreover, many emerging use cases lack established metrics for evaluating generated artifacts. For instance, in our case, determining if the LLM-generated modifications adequately address compliance requirements remains a challenge.

To address these issues, we employ a structured, three-step pipeline using an LLM to enhance compliance processes. First, we classify software use cases to determine non-compliance to specific legal provisions. Second, for use cases identified as non-compliant, we generate modifications to bring them into compliance. Third, we manually validate the proposed changes against legal standards to ensure they address the intended requirements. Our preliminary findings reveal both the potential and limitations of LLMs in this context. While the models demonstrate an ability to automate compliance tasks, their reasoning capabilities often fall short in handling nuanced legal concepts, necessitating further refinement of their frameworks.

This paper builds on prior research that explored using generative AI to evaluate software compliance with the GDPR (citation redacted). Here, we extend the focus to the CCPA, analyzing how LLMs can operationalize legal definitions of consent in a privacy context. By benchmarking LLM-generated modifications against real-world use cases, we aim to highlight the feasibility of AI-driven compliance solutions and their implications for both legal and engineering domains. Ultimately, this work contributes to bridging the gap between privacy law and software design, offering insights into the evolving role of AI in regulatory compliance.

This paper proceeds as follows: Part 2 provides the legal background on the two CCPA provisions we analyze. Part 3 provides the engineering background. Part 4 describes our research design. Part 5 discusses the principal results. Part 6 discusses both legal and engineering implications. Part 7 concludes.

% Goals for the workshop:
% * General feedback on the method
% * Specific feedback on the two provisions that we are citing (are they interesting, or are other aspects interesting?)
% * Explanation of why those provisions are good to test the method (e.g., the mid-state inference challenge)

\vspace{-1mm}

\section{Legal Background}

The California Consumer Privacy Act (CCPA), enacted in 2018 and effective in 2020, introduced robust consumer privacy protections aimed at regulating the collection, sale, and sharing of personal information by entities doing business in the state of California. In  2020, voters approved the California Privacy Rights Act (CPRA), which amends and expands the CCPA. The CPRA took effect on January 1, 2023. The CCPA, as amended by the CPRA, establishes key privacy rights for California consumers, including the right to know about personal information collected, the right to delete personal information collected from them (with some exceptions), the right to opt-out of the sale or sharing of their personal information, and the right to non-discrimination for exercising these rights. 

The CCPA, as amended by CPRA, is implemented through a combination of statutory provisions and regulations issued by the California Attorney General and the California Privacy Protection Agency (CPPA), established to enforce the CCPA and CPRA.

The law’s reach is broad, as it applies not just to businesses physically located in California but to any for-profit business that collects personal information from California residents, even if the business operates outside the state or internationally. If a business meets certain revenue or data collection thresholds and deals with California consumers' data, it is subject to the law. The CCPA covers a wide range of data types, including identifiers such as names and email addresses, IP addresses, internet browsing history, geolocation, biometric data, and more. This broad definition of personal information means that a vast array of data is subject to the CCPA’s protections. 

Central to this legal framework is the consumer’s right to opt out of businesses' sale or sharing of their personal information with third parties. \text{See Cal. Civ. Code § 1798.120 (West 2020).} The CCPA defines "sale" broadly to include any exchange of personal information for monetary or other valuable consideration, while "sharing" refers to the disclosure of personal information for targeted advertising. See id. 1798.140(ad), (ah), (k). Once a consumer exercises this right, the business must cease selling or sharing the consumer’s personal data unless the consumer provides express consent to resume those activities. The opt-out right is a key element of the CCPA/CPRA framework, giving consumers greater control over their personal data in the digital economy. 

Two special circumstances implicating the opt-out right are germane to the use cases described in this paper: (1) consumer requests to opt-in after opting-out of the sale or sharing of personal information, and (2) special rules involving the sale and sharing of minors' personal information. 

\subsection{Requests to Opt-In After Opting-Out of the Sale or Sharing of Personal
Information}
As described above, the consumer right to opt out of business' sale or sharing of their personal information with third parties is a key pillar of the CCPA/CPRA framework. But what if a consumer who has opted out to the sale or sharing of their personal information wishes to opt back in? The CCPA Regulations provide that in such an instance, "[r]equests to opt-in to the sale or sharing of personal information shall use a two-step opt-in process whereby the consumer shall first, clearly request to opt-in and then second, separately confirm their choice to opt-in." {California Consumer Privacy Act Regulations, Cal. Code Regs. tit. 11, § 7028(a).}

\subsection{Special Rules for Minors}
While the CCPA/CPRA framework allows for the sale and sharing of personal information with third parties by default unless a consumer exercises her right to opt out, the law provides a different set of default rules for minors. Instead, a business generally may \textit{not} sell or share the personal information of consumer is under 16 years old, subject to the consumer's opt-in. 
For consumers between the ages of 13 and 16, the business must obtain the consumer’s consent before selling or sharing their personal information. For these consumers, a  business "shall establish, document, and comply with a reasonable process for allowing such consumers to opt-in to the sale or sharing of their personal information." Id. § 7071.

If the consumer is under 13, the business must obtain consent from the child’s parent or guardian. See Cal. Civ. Code § 1798.120(c). The California Consumer Privacy Act Regulations provide that "[a] business that has actual knowledge that it sells or shares the personal information of a consumer less than the age of 13 shall establish, document, and comply with a reasonable method for determining that the person consenting to the sale or sharing of the personal information about the child is the parent or guardian of that child." According to the Regulations, "[m]ethods that are reasonably calculated to ensure that the person providing consent is the child’s parent or guardian include, but are not limited to:

\begin{itemize}
    \item Providing a consent form to be signed by the parent or guardian under penalty of perjury and returned to the business by postal mail, facsimile, or electronic scan;
    \item Requiring a parent or guardian, in connection with a monetary transaction, to use a credit card, debit card, or other online payment system that provides notification of each discrete transaction to the primary account holder;
    \item Having a parent or guardian call a toll-free telephone number staffed by trained personnel;
    \item Having a parent or guardian connect to trained personnel via video-conference;
    \item Having a parent or guardian communicate in person with trained personnel; and
    \item Verifying a parent or guardian’s identity by checking a form of government issued identification against databases of such information, as long as the parent or guardian’s identification is deleted by the business from its records promptly after such verification is complete.\end{itemize}
    % See California Consumer Privacy Act Regulations, Cal. Code Regs. tit. 11, § 7070.
    
\section{Engineering Background}
% * Research in generative SE, full-spectrum, focus narrowed to requirements analysis
Generative software engineering leverages AI-driven models, particularly Code-LLMs, to automate code generation, refactoring, and testing, fundamentally altering software development paradigms \cite{sarkar_what_2022}. Models like OpenAI’s CodeX \cite{chen_evaluating_2021} and DeepMind’s AlphaCode \cite{li_competition-level_2022} enable engineers to generate functions, integrate frameworks, and enhance maintainability. CodeT \cite{chen_codet_2022} automates test generation, while fine-tuned models like Coeditor assist with refactoring \cite{wei_coeditor_2024}. Open-source alternatives, such as DeepSeek-Coder \cite{guo_deepseek-coder_2024} and StarCoder \cite{li_starcoder_2023}, further democratize access to these capabilities.

Adoption has been rapid, with tools like GitHub Copilot producing approximately 35\% of users’ code \cite{thompson_how_2022}. Code-LLMs integrate with IDEs, facilitating seamless workflow enhancements \cite{xu_-ide_2021}. A key advantage is the ability to express functional requirements through declarative natural language prompts \cite{jiang_discovering_2022, shi_natural_2022}, expanding traditional software engineering paradigms. However, engineers may struggle with effective prompting, impacting model performance \cite{jiang_discovering_2022}. Hybrid approaches, such as interleaving code with declarative instructions \cite{liang_code_2023} or frameworks like R-U-SURE, which infer programmer intent \cite{johnson_r-u-sure_2023}, offer promising solutions.

Beyond efficiency, Code-LLMs enable exploratory software design, assisting in both structured development (“acceleration mode”) and problem-solving (“exploratory mode”) \cite{barke_grounded_2023}. This exploratory nature aligns with requirements engineering, where defining software constraints resembles prompt formulation for LLMs. Yang et al. (2023) use LLMs to refine requirements elicitation \cite{yang_beyond_2023}, while Ma et al. (2024) demonstrate that training on requirements enhances prompting performance \cite{ma_what_2024}.

Additionally, Code-LLMs extend into reasoning tasks beyond software engineering. Program-Aided Language (PAL) reasoning encodes logical problems as executable code, leveraging structured algorithms for enhanced reasoning performance \cite{kabra_program-aided_2023}. Formal syntax constrains problem spaces, yielding precise and computationally tractable solutions compared to natural language reasoning \cite{pan_logic-lm_2023, marra_statistical_2024}.

In this work, we leverage the generative capabilities of LLMs to automate legal compliance checking for real-world software use cases. To the best of our knowledge, we did not find any other work during our literature review that talks about applying LLMs for this requirements engineering task.

\section{Research Design} 

In software engineering, a \textit{use case} is a description of how an actor generally uses a key feature of a system.  Use cases consist of four elements: the \textit{primary actor} who initiates the use case, the \textit{pre-conditions} which must be true before a use case ``executes''; the \textit{flow of events} that describe the steps performed by the actor and the system during execution; and the \textit{post-conditions} that describe what must be true after the use case completes execution~\cite{armour_advanced_2001}. In addition to use cases, many companies use user stories written in the form of ``As an [actor], I want to [action] for the purpose of [goal]'' where the actor is a role describing a user or service, the action is what the actor aims to do, often by interfacing with the software, and the goal describes why they are performing the action. We chose to study use cases, because user stories lack the specific detail needed to verify whether regulatory requirements are being satisfied. 

In practice, we envision companies facing the question of legal compliance in two stages. First, given a large collection of use cases describing various products and systems and given a specific legal requirement in a regulation, can they answer the question, ``Which of these use cases should be inspected for compliance?'' Second, given a candidate use case for inspection, can they answer the question,``Which changes, if any, are required to bring the use case into compliance?'' In addition, while many companies develop use cases in their internal practice, they are often proprietary artifacts that are not shared publicly. Therefore, we developed a small set of use cases by hand in this study. To conduct our experiments, we use OpenAI's GPT-4o model version \texttt{gpt-4o-2024-05-13} with a cut-off date of October 23. In order to generate deterministic text using the LLM, we set the model temperature parameter to zero. 

We now discuss the principal steps in our study: 1) developing the use cases from real world product descriptions; 2) selecting use cases relevant to a particular legal requirement; and 3) modifying the use case to yield a compliant system state.

\subsection{Step 1 - Use Case Development}

%Use cases describe the flow of events that occur when a user interacts with an app functionality. A formal use case specification includes the pre-conditions to trigger the app functionality, the flow of events describing the execution of the functionality, and the post-conditions that should be met when the user interaction with the functionality is complete. These specifications serve as an important reference for software engineers to design software systems according to user requirements. 
%As explained earlier, it is necessary to factor in consent while designing user-centric software systems. Since use cases provide a bird's eye view of system design, use case specifications should include the necessary steps to ensure compliance with the mandatory legal provisions. 

\begin{table*}
\renewcommand{\arraystretch}{1.2}
\centering
\caption{Example Use Case}
\label{tab1}
\small
\begin{tabular}{ p{4cm} p{9.5cm} }
\hline
User Story &  Generated Use Case \\
\hline
As a user, I want to discover a variety of audiobooks, podcasts, and App Originals so that I can enjoy diverse storytelling. & Preconditions: 
  \begin{itemize}[leftmargin=0.2in]
     \item The user has downloaded and installed the mobile app.
      \item The user has an active account and is logged in.
      \item The app has access to the internet for content discovery.
\end{itemize} 

  Flow: 
  \begin{itemize}[leftmargin=0.2in]
      \item The user opens the mobile app.
      \item The app displays a variety of audiobooks, podcasts, and App Originals.
      \item The user navigates to the Discover section of the app.
      \item The user browses through the categories or uses the search feature to find specific content.
      \item The user selects an audiobook, podcast, or App Original to view more details.
      \item The user reads the description, reviews, and ratings of the selected content.
      \item The user adds the chosen content to their library or playlist for future listening.
  \end{itemize} 
  Postconditions: 
  \begin{itemize}[leftmargin=0.2in]
      \item The user has discovered new audiobooks, podcasts, or App Originals.
      \item The selected content is added to the user's library or playlist.
      \item The user can listen to the content immediately or later.
    \end{itemize} \\
\hline
\end{tabular}
\end{table*}

In this work, we focus on mobile applications, which have been a focus of the General Data Protection Regulation (GDPR) and CCPA in the EU and US, respectively. We first obtained a list of mobile app descriptions written by mobile app companies and published in the US Apple App and Google Play stores. These descriptions include general details about the functionality provided by the app. We next prompt the LLM to summarize a list of user stories from the mobile app description. Third, for each user story, we prompt the model to generate use cases that describe how the user story would be realized, including the pre-conditions, flow of events and post-conditions. From these lists, we hand-selected 30 use cases for experimentation. Table \ref{tab1} illustrates a user story and the corresponding use case generated from the story. 

% \vspace{-3mm}

\subsection{Step 2 - Use Case Selection}

The second step in our method involves prompting an LLM to check if the use case needs a modification to satisfy the legal provision. We treat this step as a natural language binary classification task. The LLM is expected to answer `No,' if either (1) the use case is not relevant to the provision, or (2) the use case already satisfies the provision. If either condition is false, then LLM is expected to answer `Yes.' For instance, in the case of the provision mandating a user to opt-in, we use the model to check if the use case describes any system behavior or feature that may require the sale or sharing of personal information of the user. If the model generates `No,' then we mark the use case as irrelevant and exclude it from the next stage. If the model generates `Yes,' then we mark the use case as relevant for the next stage.

We compare the answers of two prompting methods on this task. First, we use a zero-shot Yes/No prompt, wherein we instruct the model to answer a `Yes' or `No' directly. Alternately, we use zero-shot Chain-of-Thought (CoT) prompting \cite{wei_chain--thought_2023}, wherein we prompt the model to generate step-by-step reasoning before generating the Yes/No answer. Table \ref{tab2} presents the template for both prompting techniques, including the task description, answer format, and augmentation points for injecting the \textit{legal text},  \textit{app description summary}, and \textit{use case} texts shown in curly braces. The app description summary provides a one-line summary of the original mobile app description and serves as additional context to the LLM. 

% We specify a system message for the LLM stating that `it is a helpful assistant' before providing the prompt template as input.

\begin{table*}[t]
\begin{flushleft}
\renewcommand{\arraystretch}{1.2}
\centering
\caption{Prompt Templates for Use Case Selection and Modification}
\small
\label{tab2}
\begin{tabular}{ p{2.5cm} p{10cm} }
\hline
Technique &  Template \\
\hline
Zero-shot Yes/No Prompt & 
Task: Read the following legal text and mobile app description and decide if the given use case should be modified to satisfy the legal text. Respond with Yes if the use case should be modified to satisfy the legal text, or No if the use case does not need to be modified. Use JSON Format as follows: \{`Answer': `Yes / No'\} \hfill \break
Legal Text: \{\textit{legal text}\} \hfill \break
App Description Summary: \{\textit{app description summary}\} \hfill \break
Use Case: \{\textit{use case}\} \hfill \break
Answer:  \hfill \\
\hline
Zero-shot Chain-of-Thought Prompt & 
Task: Read the following legal text and mobile app description and reason step by step if the given use case should be modified to satisfy the legal text. Respond in JSON format as follows: \{`Rationale': `your rationale for the decision', `Answer': `Yes or No'\} \hfill \break
Legal Text: \{\textit{legal text}\} \hfill \break
App Description Summary: \{\textit{app description summary}\} \hfill \break
Use Case: \{\textit{use case}\} \hfill \break
Answer:  \hfill \\
\hline
Modification Prompt & 
Task: Read the following legal text, mobile app description and use case description in JSON format. Write a Python program to modify the use case description so that it satisfies the legal text. The output of the program should be a modified use case in JSON format as follows: \{`preconditions': [`condition1', `condition2', `condition3'], `flow': [`step1', `step2', `step3'], `postconditions': [`postcondition1', `postcondition2'] \}\hfill \break
Legal Text: \{\textit{legal text}\} \hfill \break
App Description Summary: \{\textit{app description summary}\} \hfill \break
Use Case: \{\textit{use case}\} \hfill \break
Answer: \\
\hline 
\end{tabular}
\end{flushleft}
\vspace{-4mm}
\end{table*}

\subsection{Step 3 - Use Case Modification}

\begin{figure}
    \centering
    \includegraphics[width=1\linewidth]{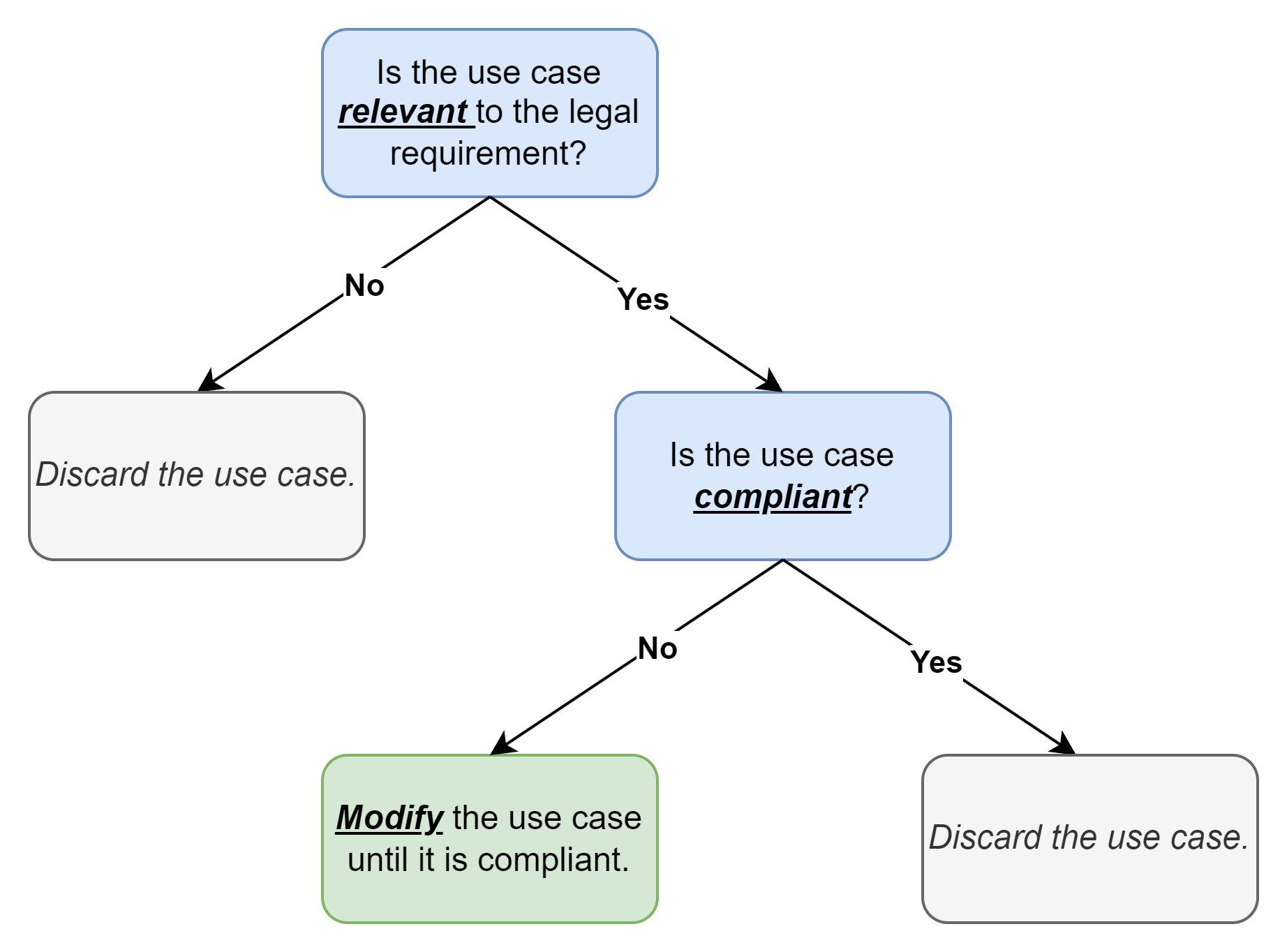}
    \caption{Criteria for use case modification}
    \label{fig:flow_chart}
    \Description{A flow chart showing the criteria used for modifying use cases.}
    \vspace{-3mm}
\end{figure}

After step two, we modify use cases that are relevant to but not compliant with a given legal requirement (see Fig. \ref{fig:flow_chart}). For each use case that needs to be modified, the model is prompted to generate a modified use case to comply with the provision. A modification may refer to the addition/deletion of at least one pre-condition, one or more steps in the flow of events, or at least one post-condition in a use case. The use cases are presented in JavaScript Object Notation (JSON) format, which is a declarative computer language for representing data. This format can be read by a computer program into memory and modified using a programming language, such as Python. By using the JSON format, we prompt the model to generate the modification in the form of a Python program that can be executed to yield the desired output. This prompting strategy implements the program-aided language (PAL) method, and also allows us to computationally tabulate results. Table \ref{tab2} shows the prompt template we used for modifying a use case.
\subsection{Research Evaluation Method}
\label{section:eval}
% * Literature and motivation for evaluation method, including human and automated techniques

We evaluate use case selection and use case modification as follows. In use case selection, a legal expert (henceforth referred to as a labeler) manually labeled the generated use cases with the correct Yes/No answer and then compute accuracy ($acc$) from the total correct answers ($tp$) and incorrect answers ($fp$) using the calculation $acc = tp / (tp + fp)$. 

For use case modification, two authors of this paper first manually review the relevant use cases for which the labeler assigned a `Yes' label to the use case. For each use case, we itemize the changes to the use case in order to satisfy the legal requirement. These changes include any one of the following operations: 
\begin{itemize}
    \item \texttt{insert\_pre(condition)}
    \item \texttt{remove\_pre(condition)}
    \item \texttt{insert\_flow(at\_index, step)}
    \item \texttt{remove\_flow(at\_index)}
    \item \texttt{insert\_post(condition)}
    \item \texttt{remove\_post(condition)}
\end{itemize}

The ground truth change list is used to compare changes with those generated by the LLM. 

%We begin by manually evaluating the modified use cases to assess their correctness against the ground truth change list. Each modification is categorized as correct, partially correct, or incorrect. A modification is labeled `correct' if it adheres to the legal provision while preserving the validity of both the pre-conditions and post-conditions of the original use case. If the modification complies with the legal provision but violates any pre-condition or post-condition, it is labeled `partially correct'. A modification that violates the legal provision is marked `incorrect'. Additionally, any change made to a use case that did not require modification is also labeled as `incorrect'.

We manually label the modified use cases using two dimensions that together define overall correctness: non-violativity, which is the labeler's determination that the change creates a state in the system that does not violate the meaning of a legal requirement; and self-consistency, which is the labeler's determination that the change is consistent with the defined pre-conditions and post-conditions. These two categorical metrics are reported separately in Section 5.2. 

In addition to manual evaluation, we also measure the syntactic similarity between the ground truth change list and the use case change list generated by the LLM. A low similarity score indicates that the sentence structure and vocabulary of the generated modifications differ significantly from the ground truth, which may suggest errors. Conversely, a high similarity score suggests that the modifications are likely correct. We calculate these syntactic similarity scores using BLEU, ROUGE-1, and ROUGE-L~\cite{papineni_bleu_2002, lin_rouge_2004}. The BLEU score assesses the precision of n-grams between the candidate and reference text, wherein an n-gram is a sequence of \textit{n} adjacent words, for example, a 2-gram is all possible word pairs~\cite{papineni_bleu_2002}. ROUGE-1 evaluates the overlap of unigrams (individual words) between the two texts~\cite{lin_rouge_2004}. ROUGE-L focuses on the longest common subsequence (LCS), emphasizing sequence similarity between the candidate and reference text~\cite{lin_rouge_2004}. 
% We also draw from a variety of techniques to automatically evaluate code generation for implementing use cases \textcolor{red}{(I think this is out of scope but trying to tie back somehow to generative SE/code gen from Engineering Background section)}. Autonomous assessment has been applied to evaluate LLM translations from formal syntax (FS) to natural language (NL) and vice versa~\cite{karia_forallutoexistsval_2024}. The idea is to utilize a formal verifier to check the fidelity of LLM translations to and from FS and NL. This approach can be applied to code generation via a compiler or interpreter to test whether synthesized programs execute properly. \textcolor{red}{program analysis and testing spiel?}

% For our study, we will use \textcolor{red}{... ??? [TODO what exactly are we evaluating? the accuracy of model predictions of use case relevance? how "well" the model generates alternative use cases that explore the search space?]}

% if a generated formal statement $\varphi'$, synthesized based on a natural language interpretation  is equivalent to its original formal statement $\varphi$. In code generation, a compiler or interpreter 

\section{Results}

\begin{table}
    \renewcommand{\arraystretch}{1.2}
    \centering
    \caption{Use Case Selection Results}
    \label{tab3}
    \small
    \begin{tabular}{c c}
       \hline
       Technique  &  Accuracy \\
       \hline
       Yes/No Prompting & 53.3\% \\
       CoT Prompting & 66.7\% \\
       \hline
    \end{tabular}
\end{table}

\begin{table}
    \centering
    \renewcommand{\arraystretch}{1.2}
    \caption{Similarity Scores for Use Case Modification}
    \small
    \label{tab4}
    \begin{tabular}{c c c c c}
    \hline
       Non-violative  & Self-Consistent & BLUE & ROUGE-1 & ROUGE-L \\
       \hline
        \checkmark & \checkmark & 0.35 & 0.68 & 0.65 \\
        \checkmark & \xmark & 0.28 & 0.51 & 0.5 \\
        \hline
    \end{tabular}
 \vspace{-3mm}   
\end{table}

\label{results}
\begin{table*}[t]
\begin{flushleft}
\renewcommand{\arraystretch}{1.2}
\centering
\small
\caption{Examples Illustrating Step 1 Model Output for the Opt-In Legal Provision}
\label{tab5}
\begin{tabular}{p{0.3cm} p{9cm} p{1cm} p{4cm} }
\hline
No. & Example Use Case &  Yes/No Prompt & CoT Prompt \\
\hline
1 &
Preconditions: 
  \begin{itemize}[leftmargin=0.2in]
     \item The user has downloaded and installed the mobile app.
      \item The user has an active account and is logged in.
      \item The app has access to the internet for content discovery.
\end{itemize} 

  Flow: 
  \begin{itemize}[leftmargin=0.2in]
      \item The user opens the mobile app.
      \item The app displays a variety of audiobooks, podcasts, and App Originals.
      \item The user navigates to the Discover section of the app.
      \item The user browses through the categories or uses the search feature to find specific content.
      \item The user selects an audiobook, podcast, or App Original to view more details.
      \item The user reads the description, reviews, and ratings of the selected content.
      \item The user adds the chosen content to their library or playlist for future listening.
  \end{itemize} 
  Postconditions: 
  \begin{itemize}[leftmargin=0.2in]
      \item The user has discovered new audiobooks, podcasts, or App Originals.
      \item The selected content is added to the user's library or playlist.
      \item The user can listen to the content immediately or later.
    \end{itemize} & 
 \textcolor{myred}{No}  & 
 
 \textcolor{mygreen}{Rationale: The legal text requires a two-step opt-in process for users who have previously opted out of the sale or sharing of their personal information. The current use case does not mention any process for handling users who have opted out and then wish to opt back in. To comply with the legal text, the use case should be modified to include a two-step opt-in process for such users, ensuring they first request to opt-in and then separately confirm their choice. \hfill \break
  Answer: Yes} \\  
\hline
2 & Preconditions:
\begin{itemize}[leftmargin=0.2in]
    \item The user has the app installed and is logged in.
    \item The user has at least one audiobook downloaded or available for streaming.
    \item The user is currently playing an audiobook.
\end{itemize}
Flow: 
\begin{itemize}[leftmargin=0.2in]
    \item The user opens the app and navigates to the audiobook currently playing.
    \item The user taps on the Sleep Timer option within the playback controls.
    \item The user selects a desired time duration for the Sleep Timer (e.g., 15, 30, or 60 minutes).
    \item The user confirms the selection of the Sleep Timer.
    \item The app sets the Sleep Timer and displays a confirmation message.
    \item The app continues playing the audiobook until the Sleep Timer expires.
    \item Once the Sleep Timer expires, the app automatically pauses the audiobook.
\end{itemize}
 Postconditions: 
 \begin{itemize}[leftmargin=0.2in]
     \item The audiobook is paused after the Sleep Timer duration expires.
     \item The user can resume listening to the audiobook later.
     \item The app remembers the user's Sleep Timer settings for future use.
 \end{itemize}
& \textcolor{mygreen}{No} & \textcolor{mygreen}{Rationale: The use case described for the mobile app's Sleep Timer feature does not involve the sale or sharing of personal information. The legal text pertains to opt-in processes for the sale or sharing of personal information, which is not relevant to the described use case. Therefore, there is no need to modify the use case to satisfy the legal text. \hfill \break
  Answer: No }  \\
  \hline
  3 & Preconditions: 
\begin{itemize}
    \item User has registered an account in the app.
    \item User has logged into the app.
    \item User has added at least one book to their reading history.
    \item User has selected their favorite genres.
    \item User has opted in to the sharing of personal information.
\end{itemize}
Flow:
\begin{itemize}
    \item User opens the app.
    \item User navigates to the 'Recommendations section.
    \item The app retrieves the user's reading history and favorite genres.
    \item The app processes the data to generate personalized book recommendations.
    \item The app displays the personalized recommendations to the user.
    \item User can browse through the recommended books.
\end{itemize}

Postconditions:
\begin{itemize}
    \item User views a list of personalized book recommendations.
    \item User can choose to read more about any recommended book.
    \end{itemize}
    & \textcolor{myred}{Yes} & \textcolor{myred} {Rationale: The legal text requires a two-step opt-in process for users who have previously opted out of the sale or sharing of their personal information. The use case does not specify whether the user had previously opted out and then opted back in using a two-step process. To comply with the legal text, the use case should be modified to ensure that if a user had previously opted out, they must go through a two-step opt-in process before their personal information can be shared for generating personalized recommendations. \hfill \break
  Answer: Yes }\\
  \hline
\end{tabular}
\end{flushleft}
\vspace{-4mm}
\end{table*}

\begin{table*}[t]
\begin{flushleft}
\renewcommand{\arraystretch}{1.2}
\centering
\small
\caption{Examples Illustrating Step 2 Model Output for the Opt-In Legal Provision}
\label{tab6}
\begin{tabular}{p{0.3cm} p{6.5cm} p{9cm} }
\hline
No. & Example Use Case &  Modification Prompt \\
\hline
1 &
Preconditions: 
  \begin{itemize}[leftmargin=0.2in]
     \item The user has downloaded and installed the mobile app.
      \item The user has an active account and is logged in.
      \item The app has access to the internet for content discovery.
\end{itemize} 

  Flow: 
  \begin{itemize}[leftmargin=0.2in]
      \item The user opens the mobile app.
      \item The app displays a variety of audiobooks, podcasts, and App Originals.
      \item The user navigates to the Discover section of the app.
      \item The user browses through the categories or uses the search feature to find specific content.
      \item The user selects an audiobook, podcast, or App Original to view more details.
      \item The user reads the description, reviews, and ratings of the selected content.
      \item The user adds the chosen content to their library or playlist for future listening.
  \end{itemize} 
  Postconditions: 
  \begin{itemize}[leftmargin=0.2in]
      \item The user has discovered new audiobooks, podcasts, or App Originals.
      \item The selected content is added to the user's library or playlist.
      \item The user can listen to the content immediately or later.
    \end{itemize} 
  & 
  Preconditions: 
  \begin{itemize}[leftmargin=0.1in]
      \item The user has downloaded and installed the mobile app.
      \item The user has an active account and is logged in.
      \item The app has access to the internet for content discovery.
      \textcolor{mygreen}{
      \item The user has previously opted out of the sale or sharing of their personal information. }
\end{itemize} 

  Flow: 
  \begin{itemize}[leftmargin=0.1in]
      \item The user opens the mobile app.
      \item The app displays a variety of audiobooks, podcasts, and App Originals.
      \item The user navigates to the Discover section of the app.
      \item The user browses through the categories or uses the search feature to find specific content.
      \item The user selects an audiobook, podcast, or App Original to view more details.
      \textcolor{mygreen}{
      \item The app informs the user that accessing this content requires the sale or sharing of their personal information and provides instructions on how to opt-in.
      \item The user clearly requests to opt-in to the sale or sharing of their personal information.
      \item The user separately confirms their choice to opt-in to the sale or sharing of their personal information.
      \item The app verifies the user's consent in compliance with section 7004.}
      \item The user reads the description, reviews, and ratings of the selected content.
      \item The user adds the chosen content to their library or playlist for future listening.
  \end{itemize} 
  Postconditions: 
  \begin{itemize}[leftmargin=0.1in]
      \item The user has discovered new audiobooks, podcasts, or App Originals.
      \item The selected content is added to the user's library or playlist.
      \item The user can listen to the content immediately or later.
      \item \textcolor{mygreen}{The user's consent to the sale or sharing of their personal information has been recorded.}
    \end{itemize} \\    
\hline
2 & 
Preconditions: 
\begin{itemize}
    \item User has registered an account in the app.
    \item User has logged into the app.
    \item User has added at least one book to their reading history.
    \item User has selected their favorite genres.
    \item User has opted in to the sharing of personal information.
\end{itemize}
Flow:
\begin{itemize}
    \item User opens the app.
    \item User navigates to the Recommendations section.
    \item The app retrieves the user's reading history and favorite genres.
    \item The app processes the data to generate personalized book recommendations.
    \item The app displays the personalized recommendations to the user.
    \item User can browse through the recommended books.
\end{itemize}

Postconditions:
\begin{itemize}
    \item User views a list of personalized book recommendations.
    \item User can choose to read more about any recommended book.
\end{itemize} &
Preconditions: 
\begin{itemize}
    \item User has registered an account in the app.
    \item User has logged into the app.
    \item User has added at least one book to their reading history.
    \item User has selected their favorite genres.
    \item User has opted in to the sharing of personal information.
\end{itemize}
Flow:
\begin{itemize}
    \item User opens the app.
    \item User navigates to the 'Recommendations section.
    \textcolor{myred}{
    \item The app checks if the user has opted in to the sharing of personal information.
    \item If the user has not opted in, the app informs the user that personalized recommendations require the sharing of personal information and provides instructions on how to opt-in.
    \item User follows the instructions to opt-in, first by clearly requesting to opt-in and then separately confirming their choice to opt-in.}
    \item The app retrieves the user's reading history and favorite genres.
    \item The app processes the data to generate personalized book recommendations.
    \item The app displays the personalized recommendations to the user.
    \item User can browse through the recommended books.
\end{itemize}

Postconditions:
\begin{itemize}
    \item User views a list of personalized book recommendations.
    \item User can choose to read more about any recommended book.
\end{itemize} \\
\hline
\end{tabular}
\end{flushleft}
\vspace{-4mm}
\end{table*}

We discuss the preliminary results of our method by checking if the use cases in our dataset comply with the opt-in legal provision. 

\subsection{Use Case Selection}

Table \ref{tab3} presents the accuracy for both Yes/No and CoT prompts in the task of use case selection. The model performs 13.4\% better with CoT prompting compared to the standard Yes/No prompt. These results indicate that when the use case selection task is framed as a binary classification problem, the LLM lacks sufficient context to predict whether a modification is needed. The improved accuracy with CoT prompting suggests that generating a rationale provides the missing context before making a final prediction.

Table \ref{tab5} provides output examples for three cases in our dataset. As highlighted in red in the table, the Yes/No prompt fails to predict the relevance of the use case to the legal provision in examples one and three. In contrast, the CoT prompt produces correct predictions for examples one and two. However, in example three, the model does not appear to attend to a key pre-condition, which states that the user has `opted in.' Since this pre-condition indicates that the user has already consented, there is no need to modify the use case to ask for consent again. Thus, the correct answer is `No,' meaning the use case should not be selected for modification.

\subsection{Use Case Modification}

Since CoT prompting performs better than standard zero-shot prompting, we use the examples flagged by the model with CoT prompting to determine the input use cases for the next stage. In our dataset of 30 use case instances, the model predicted that 12 instances required modification to comply with the legal provision. Based on the ground truth labels, eight of the 12 predictions were correct. We proceed by passing these 12 predictions as input to the next stage. We include the four false positives to observe how the model performs when presented with inaccurate information.

The manual evaluation of the modifications made by the LLM in step two showed that eight out of the 12 instances were non-violative, i.e., did not violate the legal provision. However, only two of these eight instances were self-consistent. Table \ref{tab4} compares the BLEU, ROUGE-1, and ROUGE-L scores for the instances that satisfy both the conditions of non-violativity and self-consistency against those instances that only satisfy the non-violative condition. The difference in scores between these categories indicates that use cases that satisfy both conditions are more syntactically similar to the ground truth than the ones that satisfy only one condition. 

Table \ref{tab6} demonstrates the model's output for modifying a specific use case. As shown in the table, the LLM successfully modifies the input use case in example one, following the instructions provided in the modification prompt. However, in example two, the LLM introduces unnecessary steps for opt-in, despite the pre-condition already indicating the user's consent preference. Although this modification is incorrect, it is understandable given the faulty prediction made in the previous stage of the pipeline (i.e., the lack of attention to critical pre-conditions).

% \vspace{-3mm}

\section{Implications}

\subsection{Legal Implications}

These tentative results lead to some early lessons for the development of privacy law in both Europe and U.S. jurisdictions enacting privacy legislation. One major lesson from our investigation is that compliance remains a major challenge for privacy enforcement. Much of the legal literature focuses on the enforcement gap created by information asymmetries between firms and regulators. We shed light on another dimension of this problem - \textit{how} to comply can be confusing even for good faith actors. Even legislative provisions that seem clearly written and specific raise many questions surrounding how to bridge the gap between the legal principle and engineering practice.

One potential implication here is that the role of administrative agencies such as the CPPA will have a big role to play in making sure that legislation becomes effective and informative. While there will almost certainly be litigation to clarify some of the unclear aspects of the two provisions we studied alongside others, this process can be slow and not necessarily comprehensive. Drafting regulations and guidance that give clear rules of the road to engineers can help bridge much of the gap we see in our study.

Second, our investigation suggests that both the consent and opt-in/opt-out frameworks underlying CCPA/CPRA, and GDPR, should be studied more empirically. Much of the empirical literature on these questions have surrounded whether consumers understand and exercise their rights. Comparatively little attention has been paid to how these mechanisms actually get implemented in code. An important prerequisite for consumers exercising control over their data is that they are actually given meaningful opportunities to do so. When developers are unclear on how to provide such options, it becomes difficult to assess whether these provisions are working as intended from the consumers' perspective.

Third, we suggest that as software development becomes increasingly generative in the future, law will need to contemplate this development. As generative AI is used to generate software requirements and code, how generative AI understands legal requirements will be an important predictor of whether that software is ultimately compliant. Lawmakers should pay attention to these dynamics, and consider how to promote training generative AI that prioritizes compliance with laws like CCPA/CPRA. Of course, this insight has broader implications beyond privacy law and should be considered for any area of law that gets implemented in code.

Looking ahead to implications for privacy law scholarship generally, we plan to expand this methodology to studies of European privacy law, as well as other jurisdictions. Questions surrounding how to operationalize consent are even more complicated in some other contexts than what we have started with in this study. For example, the General Data Protection Regulation requires that consent be ``freely given, specific, informed and unambiguous.'' How should developers thinks about implementing this kind of standard in code? Such a standard is less specific than the two CCPA provisions we study, yet represents a major compliance task for any firm with a digital presence in Europe.

This question about the GDPR shows how these results will have implications for the privacy law community broadly. First, we will contribute a valuable empirical assessment of how ``consent'' is reflected by generative language models to those who need to implement consent by design. Much discourse has focused on whether consumers understand consent requirements, but comparatively little has addressed how businesses and software engineers understand them, and even less the coming revolution in large language models generating requirements. Second, the empirical methodology leverages advances in natural language processing that can be adapted to study privacy concepts in other textual sources such as privacy policies and corporate disclosures. Third, our results can help indicate whether regulatory interventions are better geared toward enhancing the current choice regime, or whether the regime is unworkable and other interventions should be explored.

% \vspace{-3mm}

\subsection{Engineering Implications}

Based on the preliminary results discussed in Section \ref{results}, we can draw the following implications:

1) \textbf{Reasoning Limitations in LLMs:} As demonstrated in Section \ref{results}, LLMs frequently struggle to reason accurately in our specific task. This highlights the need to avoid relying solely on their generative capabilities for legal analysis of use cases. Furthermore, LLMs are prone to hallucinations when given complex queries as input which can affect the correctness of the output. To address these limitations, we plan to work on formally representing the required reasoning patterns using neurosymbolic methods and supplying these formal representations as external context to the LLM. By augmenting formal reasoning patterns with LLMs' generative prowess, we aim to improve the overall accuracy of use case selection and mitigate the occurrence of false negatives and false positives. 

Furthermore, given accuracy results that fall below 70\%, future approaches should avoid deciding whether a system is compliant, which is a consequential decision that necessitates experience and training in law and the ability to incorporate a broader legal context. Instead, engineering solutions should focus on identifying evidence of non-compliant behavior in software systems for further review by legal experts. 

2) \textbf{Quantitative Evaluation of Use Case Modification:} While the use case modification prompt helps in introducing additional pre-conditions, flow steps, and post-conditions to ensure compliance with a legal provision, we rely on manual validation by subject matter experts (SMEs) to assess the violativity and self-consistency of the software artifacts. Going forward, we plan to incorporate unit testing for the automated evaluation of modified use cases. Using a set of legally compliant use cases as a reference, we will derive heuristics based on different ways of modifying a use case, and then use these heuristics to create individual test cases.

Additionally, we aim to evaluate the robustness of use case exploration. The main objective of our approach is to facilitate thorough exploration of legal design decisions for software. The modified use cases illustrate some capacity for the LLM to propose changes, yet exploration should sample from a larger number of changes to provide engineers more options. Yang et al. (2023) operationalize ``comprehensive exploration'' by clustering generated statements and then using the number of distinct ``concept clusters'' covered by the LLM output. These concept clusters serve as a proxy measurement of exploration and are created using SentenceBERT and Hierarchical Clustering~\cite{yang_beyond_2023, ward_hierarchical_1963}.

Lastly, we acknowledge the limitations of automated metrics like BLEU and ROUGE, introduced in Section \ref{section:eval}. These limitations are discussed at length by many others~\cite{zhang_rouge-sem_2024, akter_revisiting_2022,cohan_revisiting_2016, callison-burch_re-evaluating_2006, nenkova_summarization_2006}. When evaluating generated text, BLEU and ROUGE rely on exact-match heuristics at the n-gram level rather than holistic semantic approaches that consider broader context. As a result, these metrics may fail to account for synonyms or different word tenses when assessing similarity. They may miss the overall meaning and broader context of a sentence, which a human presumably uses to evaluate quality, due to their narrower approach at the n-gram level. 

When used to evaluate code generation, these metrics face additional shortfalls. \citet{evtikhiev_out_2023} highlight the lack of understanding about aligning automated metrics with human judgment, specifically for code generation. To mitigate these limitations, we foresee a opportunities to explore task decomposition in which we prompt the model to analyze elements of a use case independently where semantic measures, such as Cosine similarity, may be used.

3) \textbf{Qualitative Evaluation of Use Cases:} Subject matter expertise for this problem is distributed across two professions: engineers are likely to best understand self-consistency, especially when technology is involved; and lawyers are likely to best understand violativity, including the how compliance and legal violation are understand in the context of untested law. To study improve upon qualitative metrics, we must unpack these concepts within each profession to develop better heuristics and sources of information needed to collect SME judgements. The subjective nature of legal interpretation requires nuanced assessment by legal specialists. How legal analysts frame compliance questions is a potential starting point. In contrast, for use case \textit{modification}, we will explore the role software analysts in making these decisions.

% \vspace{-2mm}

\section{Conclusion}

This paper examines how consent requirements under the CCPA are understood and implemented in software systems, with a focus on the impact of generative AI. To operationalize this problem, we focused on analyzing the legal compliance of Software Engineering use cases. We incorporated an automated pipeline based on prompting Large Language Models (LLMs) to modify non-compliant use cases. We evaluated the modified use cases using automated metrics and also manually validated individual instances in our dataset. Our results highlight that while AI tools can assist in interpreting legal definitions of consent, they also present challenges, such as inaccuracies and potential misunderstandings of complex legal obligations. The ability of AI to model these requirements effectively is critical to ensuring compliance, but current generative AI models lack the ability to capture the nuances of consent requirements effectively. In the future, our goal is to devise alternate prompting methods to address the limited reasoning capabilities of generative AI tools on this task. We also plan to devise automated evaluation strategies to enable stakeholders to validate model output.

\bibliographystyle{ACM-Reference-Format}  % or whatever ACM style you prefer
\bibliography{references}              % .bib file without extension

% \printbibliography

\end{document}